# Nanodevices and Maxwell's Demon

*Supriyo Datta*
*School of Electrical and Computer Engineering,*
*NSF Network for Computational Nanotechnology,*
*Purdue Universtiy, West Lafayette, IN-47906, USA*.

**(Date: April 12, 2007)**

In the last twenty years there has been significant progress in our understanding of quantum transport far from equilibrium and a conceptual framework has emerged through a combination of the Landauer approach with the non-equilibrium Green function (NEGF) method, which is now being widely used in the analysis and design of nanoscale devices. It provides a unified description for all kinds of devices from molecular conductors to carbon nanotubes to silicon transistors covering different transport regimes from the ballistic to the diffusive limit. In this talk I use a simple version of this model to analyze a specially designed device that could be called an electronic Maxwell's demon, one that lets electrons go preferentially in one direction over another. My objective is to illustrate the fundamental role of "contacts" and "demons" in transport and energy conversion. The discussion is kept at an academic level steering clear of real world details, but the illustrative devices we use are very much within the capabilities of present-day technology. For example, recent experiments on thermoelectric effects in molecular conductors agree well with the predictions from our model. The Maxwell's demon device itself is very similar to the pentalayer spin-torque device which has been studied by a number of groups though we are not aware of any discussion of the possibility of using the device as a nanoscale heat engine or as a refrigerator as proposed here. However, my objective is not to evaluate possible practical applications. Rather it is to introduce a simple transparent model showing how out-of-equibrium demons suitably incorporated into nanodevices can achieve energy conversion.

## 1. INTRODUCTION

Maxwell invented his famous demon to illustrate the subtleties of the second law of thermodynamics and his conjecture has inspired much discussion ever since [see for example, Leff and Rex,1990,2003, Nikulov and Sheehan, 2004]. When the subject of thermodynamics was relatively new, it was not clear that heat was a form of energy since heat could never be converted entirely into useful work. Indeed the second law asserts that none of it can be converted to work if it is all available at a single temperature. Heat engines can only function by operating between two reservoirs at two different temperatures. Maxwell's demon is supposed to get around this fundamental principle by creating a temperature differential between two sides of a reservoir that is initially at a uniform temperature. This is achieved by opening and closing a little door separating the two sides at just the right times to allow fast molecules (white) to cross to the left but not the slow molecules (black, Fig.1a). As a result, faster molecules crowd onto the left side making its temperature higher than that of the right side.

Technology has now reached a point where one can think of building an electronic Maxwell's demon that can be interposed between the two contacts (labeled source and drain) of a nanoscale conductor (Fig.1b) to allow electrons to flow preferentially in one direction so that a current will flow in the external circuit even without any external source of power. Such a device can be built (indeed one could argue has already been built) though not surprisingly it is expected to operate in conformity with the second law. Here I would like to use this device simply to illustrate the fundamental role of "contacts" and "demons" in transport and energy conversion. I will try to keep the discussion at an academic level steering clear of real world details. But it should be noted that the illustrative devices we will discuss are very much within the capabilities of present-day technology. For example, recent pioneering experiments on thermoelectric effects in molecular conductors [Reddy et.al.2007] seem to agree well with the predictions from our model [Paulsson and Datta,2003]. The device discussed in Section 3 (Figs.5,6) incorporating an elastic Maxwell's demon (Datta 2005b,c) is being investigated experimentally by introducing manganese impurities into a GaAs spin-valve device with MnAs contacts [Saha et.al.2006]. The device incorporating an inelastic demon described in Section 4 (Figs.9,10) is very similar to the pentalayer spin-torque device which has been studied by a number of groups both experimentally [Fuchs et.al.2006,

**(a) Maxwell's demon**     **(b) Electronic demon**

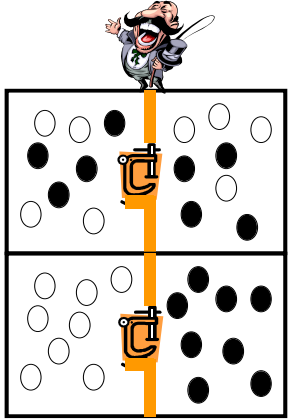
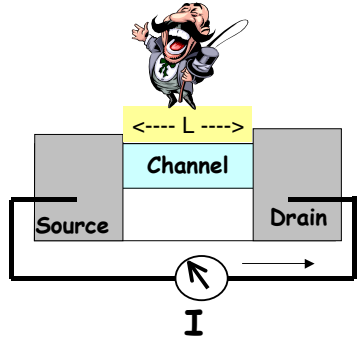

*Fig.1. (a) Maxwell's demon opens and closes a trapdoor to separate fast (white) molecules from slow (black) molecules making the left warmer than the right, thus creating a temperature differential without expending energy . (b) Electronic Maxwell's demon discussed in this talk lets electrons go preferentially from right to left thus creating a current in the external circuit without any external source of energy.*

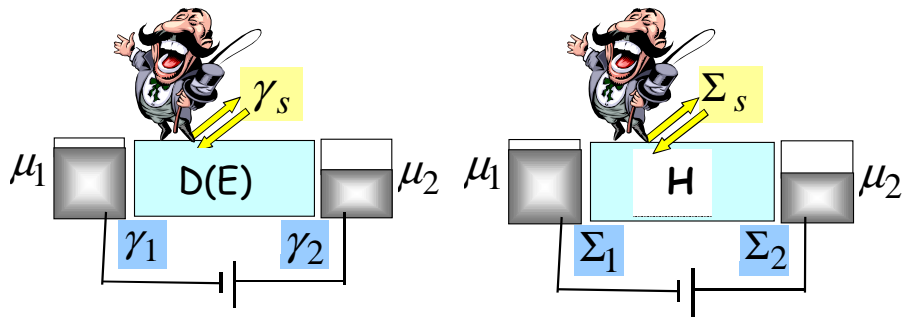

*Fig.2. Schematic representing the general approach used to model nanoscale devices: (a) Simple version with numbers $\gamma$, D used in this talk and (b) Complete version with matrices $\Sigma$,H (Adapted from Datta2005a).*

*Supriyo Datta*

Huai et.al.2004] and theoretically [Salahuddin and Datta,2006a,b] though we are not aware of any discussion of the possibility of using the device as a nanoscale heat engine or as a refrigerator as proposed here. We leave it to future work to assess whether these possibilities are of any practical importance. Here my objective is to lay out a simple transparent model showing how out-of-equilibrium demons suitably incorporated into nanodevices can achieve energy conversion.

In the last twenty years there has been significant progress in our ability to tackle the problem of quantum transport far from equilibrium and a conceptual framework has emerged through a combination of the Landauer approach with the non-equilibrium Green function (NEGF) method [Datta,1989,1990, Meir and Wingreen,1992],which is being widely used in the analysis and design of nanoscale devices [see Datta,2005a and references therein]. It provides a unified description for all kinds of devices from molecular conductors to carbon nanotubes to silicon transistors in terms of the Hamiltonian [H] describing the channel, the self-energies [$\Sigma_{1,2}$] describing the connection to the contacts and the [$\Sigma_s$] describing interactions inside the channel (Fig.2b). In each case the details are very different, but once these matrices (whose size depends on the number of basis functions needed to describe the channel) have been written down, the procedure for obtaining quantities of interest such as current flow and energy dissipation are the same regardless of the specifics of the problem at hand. The model covers different transport regimes from the ballistic to the diffusive limit depending on the relative magnitudes of $\Sigma_{1,2}$ and $\Sigma_s$. In this paper I will use a particularly simple version of this approach (Fig.2a) where matrices like $\Sigma_{1,2}$ and $\Sigma_s$ are replaced with numbers like $\gamma_{1,2}$ and $\gamma_s$ having simple physical interpretations: $\gamma_{1,2}/\hbar$ represents the rate of transfer of channel electrons in and out of the contacts while $\gamma_s/\hbar$ represents the rate at which they interact with any "demons" that inhabit the channel.

In the past it was common to have $\gamma_{1,2} \ll \gamma_s$ so that transport was dominated by the interactions within the channel, with contacts playing a minor enough role that theorists seldom drew them prior to 1990! By contrast, today's nanodevices have reached a point where $\gamma_{1,2} \gg \gamma_s$, placing them in what we could call the ballistic or "Landauer limit". An appealing feature of this limit is that *it physically separates the dynamics from the dissipation:* reversible dynamics dominates the channel while dissipation dominates the contact. Usually these two aspects of transport are conceptually so intertwined that it is difficult to see how irreversibility is introduced into a problem described by reversible dynamic equations (Newton or Schrodinger) and this issue continues to spark debate and discussion ever since the path-breaking work of Boltzmann many years ago [see for example, McQuarrie,1976].

Let me start with a brief summary of the basic framework shown in Fig.2a that we call the bottom-up viewpoint (Section 2). We will then use this approach to discuss a specially designed device which is in the Landauer limit except for a particularly simple version of Maxwell's demon that interacts with the channel electrons but does not exchange any energy with them (Section 3).We then consider a more sophisticated demon that exchanges energy as well and show how it can be used to build nanoscale heat engines and refrigerators (Section 4). We conclude with a few words about entangled demons and related conceptual issues that I believe need to be clarified in order to take transport physics to its next level (Section 5). Readers may find the related video lectures posted on the nanoHUB useful [Datta,2006] and I will be happy to share the MATLAB codes used to generate the figures in this paper.

## 2. BOTTOM-UP VIEWPOINT

Consider the device shown in Fig.1b without the "demon" but with a voltage *V* applied across two contacts (labeled "source" and "drain") made to a conductor ("channel"). How do we calculate the current *I*, as the length of the channel *L* is made shorter and shorter, down to a few atoms? This is not just an academic question since experimentalists are actually making current measurements through "channels" that are only a few atoms long. Indeed, this is also a question of great interest from an applied point of view, since every laptop computer contains about one billion nanotransistors, each of which is basically a conductor like the one in Figure 1b with L ~ 50 nm, but with the demon replaced by a third terminal that can be used to control the resistance of the channel.

As the channel length *L* is reduced from macroscopic dimensions (~ millimeters) to atomic dimensions (~ nanometers), the nature of electron transport that is, current flow, changes significantly (Fig.3). At one end, it is described by a diffusion equation in which electrons are viewed as particles that are repeatedly scattered by various obstacles causing them to perform a "random walk" from the source to the drain. The resistance obeys Ohm's law: a sample twice as long has twice the resistance. At the other end, there is the regime of ballistic transport where the resistance of a sample can be independent of length. Indeed due to wavelike interference effects it is even possible for a longer sample to have a lower resistance!

The subject of current flow is commonly approached using a "top-down" viewpoint. Students start in high school from the macroscopic limit (large *L*) and seldom reach the atomic limit, except late in graduate school if at all. I believe that this is primarily for historical reasons. After all, twenty years ago, no one knew what the resistance was for an atomic scale conductor, or if it even made sense to

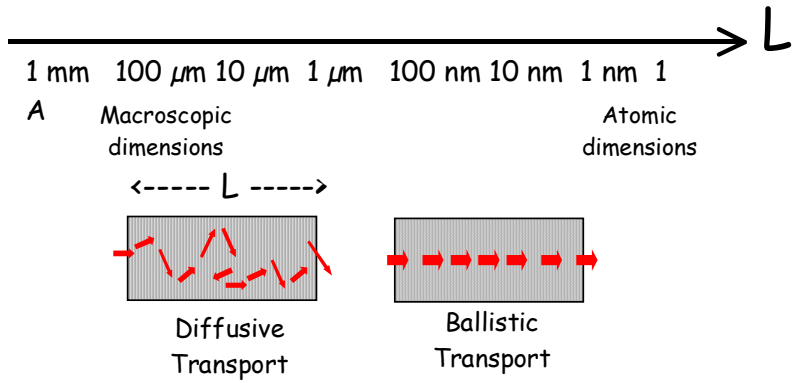

*Fig.3. Evolution of devices from the regime of diffusive transport to ballistic transport as the channel length L is scaled down from millimeters to nanometers.*

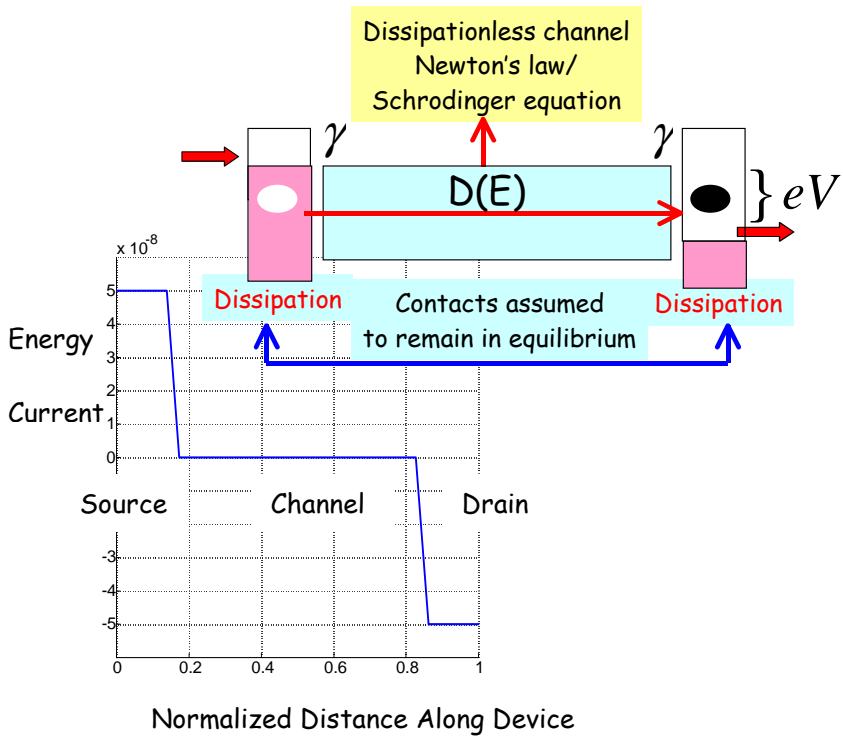

*Fig.4. Plot of energy current in a 1-D ballistic conductor with $G = e^2/h \approx 40\,\mu S$ and an applied voltage of $V = 0.05$ volts. Energy dissipated is given by the drop in the energy current, showing that the Joule heating $V^2 G = 0.1\,\mu W$ is divided equally between the two contact-channel interfaces.*

*Supriyo Datta*

ask about its resistance But now that the bottom-line is known, I believe that a "bottom-up" approach is needed if only because nanoscale devices like the ones I want to talk about look too complicated from the "top-down" viewpoint.

In the top-down view we start by learning that the conductance G=I/V is related to a material property called conductivity $\sigma$ through a relation that depends on the sample geometry and for a rectangular conductor of cross-section A and length L is given by $G = \sigma A / L$. We then learn that the conductivity is given by

$$\sigma = e^2 n \tau / m$$

where e is the electronic charge, n is the electron density, $\tau$ is the mean free time and m is the electron mass. Unfortunately from this point of view it is very difficult to understand the ballistic limit. Since electrons get from one contact to the other without scattering it is not clear what the mean free time $\tau$ is. Neither is it clear what one should use for 'n' since it stands for the density of free electrons and with molecular scale conductors it may not be clear which electrons are free. Even the mass is not very clear since the effective mass is deduced from the bandstructure of an infinite periodic solid and cannot be defined for really small conductors.

**2.1. Conductance formula: the "bottom-up" version**

A more transparent approach at least for small conductors is provided by the bottom-up viewpoint [Datta, 2005a, Chapter 1] which leads to an expression for conductance in terms of two basic factors, namely the density of states D around the equilibrium electrochemical potential and the effective escape rate $\gamma/\hbar$ from the channel into the contacts ($\hbar = h/2\pi$, h being Planck's constant):

$$G = (e^2/h) 2\pi D \gamma \qquad (1a)$$

The escape rate appearing above is the series combination of the escape rates associated with each contact:

$$\gamma = \gamma_1 \gamma_2 / \gamma_1 + \gamma_2 \qquad (1b)$$

This is an expression that we can apply to the smallest of conductors, even a hydrogen molecule. Although it looks very different from the expression for conductivity mentioned earlier, it is closer in spirit to another well-known expression for the conductivity

$$\sigma = e^2 \tilde{N} \tilde{D} \qquad (2)$$

in terms of the density of states per unit volume $\tilde{N}$ and the diffusion coefficient $\tilde{D}$. Indeed we could obtain Eq.(1a) from Eq.(2) if we make the replacements

$$D \rightarrow \tilde{N}.AL \quad \text{and} \quad \gamma/\hbar \rightarrow \tilde{D}/L^2 \qquad (3)$$

which look reasonable since the density of states for a large conductor is expected to be proportional to the volume AL and the time taken to escape from a diffusive channel is ~ $\tilde{D}/L^2$.

**2.2 Current-voltage relation: without demons**

The result cited above (Eq.(1a)) is a linear response version of a more general set of equations that can be used to calculate the full current-voltage characteristics, which in turn follow from the NEGF-Landauer formulation (Fig.2b). For our purpose in this talk the simpler version (Fig.2a) will be adequate and in this version the basic equations are fairly intuitive:

$$\begin{aligned} i_1(E) &= (e/\hbar)\, \gamma_1\, D(E)\,[f_1(E)-f(E)] \\ i_2(E) &= (e/\hbar)\, \gamma_2\, D(E)\,[f(E)-f_2(E)] \end{aligned} \qquad (4)$$

These equations relate the currents per unit energy at contacts 1 and 2 to the density of states D(E), the Fermi functions at the two contacts related to their electrochemical potentials

$$f_{1,2}(E) = \frac{1}{1+\exp((E-\mu_{1,2})/k_B T_{1,2})} \qquad (5)$$

and the distribution function f(E) inside the channel.

The total currents at the source and drain contacts are obtained by integrating the corresponding energy resolved currents

$$I_1 = \int dE\, i_1(E) \quad , \quad I_2 = \int dE\, i_2(E) \qquad (6)$$

If the electrons in the channel do not interact with any demons, we can simply set $i_1(E) = i_2(E)$, calculate f(E) and substitute back into Eq.(4) to obtain

$$\begin{aligned} i_1(E) &= i_2(E) \\ &= (e/\hbar)\frac{\gamma_1 \gamma_2}{\gamma_1+\gamma_2} D(E)\,[f_1(E)-f_2(E)] \end{aligned} \qquad (7)$$

The conductance expression stated earlier (Eq.(1a)) follows from Eqs.(6) and (7) by noting that

$$\int dE\,[f_1(E)-f_2(E)] = \mu_1 - \mu_2 \qquad (8)$$

and assuming the density of states D(E) to be nearly constant over the energy range of transport where $f_1(E) - f_2(E)$ is significantly different from zero.

***One-level conductor:*** Eq.(7) can be used more generally even when the density of states has sharp structures in energy. For example, for a very small conductor with just one energy level in the energy range of interest, the density of states is a "delta function" that is infinitely large at a particular energy. Eqs.(6,7) then yield a current-voltage characteristic as sketched below.

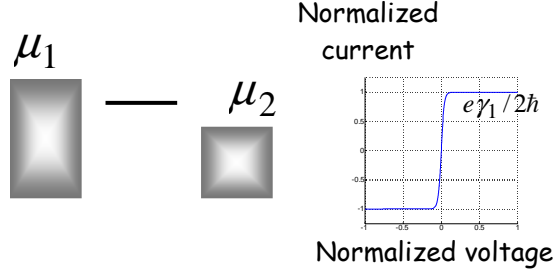

The maximum current is equal to $e\gamma_1/2\hbar$, assuming $\gamma_2 = \gamma_1$. It might appear that the maximum conductance can be infinitely large, since the voltage scale over which the current rises is $\sim k_B T$, so that dI/dV can increase without limit as the temperature tends to zero. However, the uncertainty principle requires that ***the escape rate of $\gamma/\hbar$ into the contacts from an energy level also broadens the level by $\gamma$*** as shown below. This means that the voltage scale over which the current rises is at least $\sim (\gamma_1 + \gamma_2)/e = 2\gamma_1/e$, even at zero temperature.

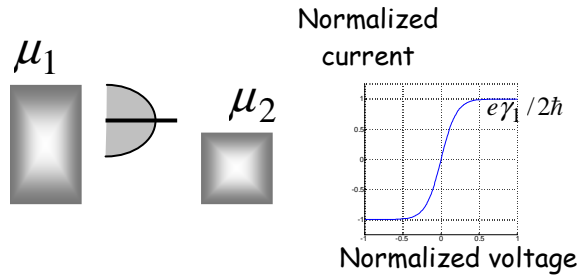

This means that a small device has a maximum conductance of

$$\frac{dI}{dV} \sim \frac{e\gamma_1/2\hbar}{2\gamma_1} = \frac{e^2}{4\hbar}$$

This rough estimate is not too far from the correct result

$$G_{max} = e^2/h \approx 25.8\ K\Omega \qquad (9)$$

which is one of the seminal results of mesoscopic physics that was not known before 1988. One could view this as a consequence of the energy broadening required by the uncertainty principle which comes out automatically in the full Landauer-NEGF approach (Fig.2b), but has to be inserted by hand into the simpler version we are using where the density of states D(E) is an input parameter (Fig.2a).

For our examples in this paper we will use a density of states that is constant in the energy range of interest, for which the current-voltage characteristic is basically linear.

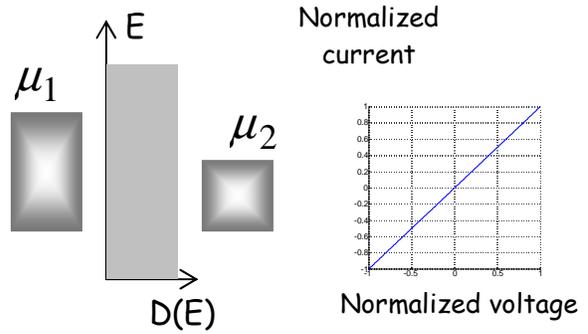

### 2.3 Current-voltage relation: with demons

Defining $i_s(E)$ as the "scattering current" ***induced by interaction of the electrons with the "demon"***, we can write

$$i_1(E) = i_2(E) + i_s(E) \qquad (10)$$

This current can be modeled in general as a difference between two processes one involving a loss of energy $\varepsilon$ from the demon and the other involving a gain of energy by the demon.

$$\begin{aligned} i_s(E) = \ & (e/\hbar)\, \gamma_s\, D(E) \int d\varepsilon\, D(E+\varepsilon) \\ & [F(\varepsilon) f(E)(1 - f(E+\varepsilon)) \\ & - F(-\varepsilon) f(E+\varepsilon)(1 - f(E))] \end{aligned} \qquad (11)$$

The basic principle of equilibrium statistical mechanics requires that if the demon is in equilibrium at some temperature $T_D$ then the strength $F(\varepsilon)$ of the energy loss processes is related to the strength $F(-\varepsilon)$ of energy gain processes by the ratio:

$$F(\varepsilon) = F(-\varepsilon) \exp(-\varepsilon/k_B T_D) \qquad (12)$$

With a little algebra one can show that this relation ensures that if the electron distribution f(E) is given by an equilibrium Fermi function with the same temperature T the two terms in Eq.(11) will cancel out. This result is independent of the detailed shape of the function $F(\varepsilon)$ describing the spectrum of the demon, as long as Eq.(12) is true. This means that *if the demon is in equilibrium with the electrons with the same temperature, there can be no net flow of energy either to or from the demon*. Indeed one could view this as the basic principle of equilibrium statistical mechanics and work backwards to obtain Eq.(12) as the condition needed to ensure compliance with this principle.

To summarize, if the electrons in the channel do not interact with any "demons", the current voltage characteristics are obtained from Eq.(7) using the Fermi functions from Eq.(5). For the more interesting case with interactions, we solve for the distribution f(E) inside the channel from Eqs.(4),(10) and (11) and then calculate the currents. Usually the current flow is driven by an external voltage that separates the electrochemical potentials $\mu_1$ and $\mu_2$ in Eq.(7). But thermoelectric currents driven by a difference in temperatures $T_1$ and $T_2$ can also be calculated from this model [Paulsson and Datta,2003] as mentioned in the introduction. Our focus here is on a different possibility for energy conversion, namely through out-of-equilibrium demons.

**2.4. Where is the heat?**

Before we move on, let me say a few words about an important conceptual issue that caused much argument in the early days: *Where is the heat dissipated in a ballistic conductor?* After all, if there is no demon to take up the energy, there cannot be any dissipation inside the channel. The answer is that the transiting electron appears as a hot electron in the drain (right) contact and leaves behind a hot hole in the source (left) contact (see Fig.4). The contacts are immediately restored to their equilibrium states by unspecified dissipative processes operative within each contact. These processes can be quite complicated but are usually incorporated surreptitiously into the model through what appears to be an innocent boundary condition, namely that the electrons in the contacts are always maintained in thermal equilibrium described by Fermi distributions (Eq.(5)) with electrochemical potentials $\mu_1$ and $\mu_2$ and temperatures $T_1$ and $T_2$.

To understand the spatial distribution of the dissipated energy it is useful to look at the energy current which is obtained by replacing the charge 'e' with the energy E of the electron:

$$I_{E1} = (1/e) \int dE \, E \, i_1(E) \quad , \qquad (13)$$
$$I_{E2} = (1/e) \int dE \, E \, i_2(E)$$

The energy currents at the source and drain contacts are written simply as

$$I_{E_s}(E) = (\mu_1/e) I_1 \quad , \qquad (14)$$
$$I_{E_d}(E) = (\mu_2/e) I_2$$

assuming that the entire current flows around the corresponding electrochemical potentials,

Fig.4 shows a spatial plot of the energy current from the source end to the drain end for a uniform 1-D ballistic conductor with a voltage of 50 mV applied across it. For a conductor with no demon for electrons to exchange energy with, $i_1(E) = i_2(E)$ making the energy current uniform across the entire channel implying that no energy is dissipated inside the channel. But the energy current entering the source contact is larger than this value while that leaving the drain contact is lower. Wherever the energy current drops, it means that the rate at which energy flows in is greater than the rate at which it flows out, indicating a net energy dissipation. Clearly in this example, 0.05 μW is dissipated in each of the two contacts thus accounting for the expected Joule heating given by $V^2 G = 0.1$ μW.

Real conductors have distributed demons throughout the channel so that dissipation occurs not just in the contacts but in the channel as well. Indeed we commonly assume the Joule heating to occur uniformly across a conductor. But there are now experimental examples of nanoscale conductors that would have been destroyed if all the heat were dissipated internally and it is believed that the conductors survive only because most of the heat is dissipated in the contacts which are large enough to get rid of it efficiently. The idealized model depicted in Fig.4 thus may not be too far from real nanodevices of today.

As I mentioned in the introduction, what distinguishes the Landauer model (Fig.4) is the physical separation of dynamics and dissipation clearly showing that what makes transport an irreversible process is the continual restoration of the contacts back to equilibrium. Without this *repeated restoration*, all flow would cease once a sufficient number of electrons have transferred from the source to the drain. Over a hundred years ago Boltzmann showed how pure Newtonian dynamics could be supplemented to describe transport processes, and his celebrated equation stands today as the centerpiece for the flow of all kinds of particles. Boltzmann's approach too involved "repeated restoration" through an assumption referred to as the "Stohsslansatz" [see for example, McQuarrie 1976] Today's devices often involve Schrodinger dynamics in place of Newtonian dynamics and the non-equilibrium Green function (NEGF) method that we use (Fig.2b) supplements the Schrodinger equation with similar assumptions about the repeated restoration of the surroundings that enter the evaluation of the various self-energy functions $\Sigma$ or the corresponding quantities $\gamma$ in the simpler model that we are using.

"Contacts" and "demons" are an integral part of all devices, the most common demon being the phonon bath for which the relation in Eq.(12) is satisfied by requiring that energy loss ~ N and energy gain ~ (N+1), N being the number of phonons given by the Bose Einstein factor if the bath is maintained in equlibrium. Typically such demons add channels for dissipation, but our purpose here is to show how suitably engineered out-of-equilibrium demons can act as sources of energy.

For this purpose, it is convenient to study a device specially designed to accentuate the impact of the demon. Usually the interactions with the demon do not have any clear distinctive effect on the terminal current that can be easily detected. But in this special device, ideally no current flows *unless the channel electrons interact with the demon*. Let me now describe how such a device can be engineered.

## 3. ELASTIC DEMON

Let us start with a simple 1-D ballistic device but having two rather special kinds of contacts. The source contact allows one type of spin, say "black" (upspins, drawn pointing to the right), to go in and out of the channel much more easily than the other type, say "white" (downspins, drawn pointing to the left, Fig.5a). Devices like this are called spin-valves and are widely used to detect magnetic fields in magnetic memory devices [see for example the articles in Heinreich and Bland, 2004].

Although today's spin-valves operate with contacts that are far from perfect, since we are only trying to make a conceptual point, let us simplify things by assuming contacts that are perfect in their discrimination between the two spins. One only lets black spins while the other only lets white spins to go in and out of the channel. We then have the situation shown in Fig.5b and no current can flow since neither black nor white spins communicate with both contacts. But if we introduce impurities into the channel (the demon) with which electrons can interact and flip their spin, then current flow should be possible as indicated: black spins come in from the left, interact with impurities to flip into white spins and go out the right contact (Fig.5c).

Consider now what happens if the impurities are say all white (Fig.6a). Electrons can now flow only when the bias is such that the source injects and drain collects (positive drain voltage), but not if the drain injects and the source collects (negative drain voltage). This is because the source injects black spins which interact with the white impurities, flip into whites and exit through the drain. But the drain only injects white spins which cannot interact with the white impurities and cannot cross over into the source. Similarly if the impurities are all black, current flows only for negative drain voltage (Fig.6b). At non-zero temperatures the cusps in the current-voltage characteristics get smoothed out and we get the smoother curves shown in Figs.6a,b.

Note the surprising feature of the plots at T=300K: *there is a non-zero current even at zero voltage!* This I believe is correct. Devices like those shown in Figs.6a,b with polarized impurities could indeed be used to generate power and one could view the system of impurities as a Maxwell's demon that lets electrons go preferentially from source to drain or from drain to source. The second law is in no danger, since the energy is extracted at the expense of the entropy of the system of impurities that collectively constitute the demon. Assuming the spins are all non-interacting and it takes no energy to flip one, the polarized state of the demon represents an unnatural low entropy state. Every time an electron goes through, a spin gets flipped raising the entropy and the flow will eventually stop when demon has been restored to its natural unpolarized state with the highest entropy of $N_I k_B \ln 2$, where $N_I$ is the number of impurities. To perpetuate the flow an external source will have to spend the energy needed to maintain the demon in its low entropy state.

### 3.1. Current versus voltage: Model summary

Let me now summarize the equations that I am using to analyze structures like the one in Fig.6 quantitatively. Basically it is an extension of the equations described earlier (see Eq.(4), Section 2) to include the two spin channels denoted by the subscripts 'u' (up or black) and 'd' (down or white):

$$i_{1,u}(E) = (e/\hbar)\, \gamma_{1,u}\, D_u(E)\, [f_1(E) - f_u(E)]$$
$$i_{1,d}(E) = (e/\hbar)\, \gamma_{1,d}\, D_d(E)\, [f_1(E) - f_d(E)] \quad (15)$$
$$i_{2,d}(E) = (e/\hbar)\, \gamma_{2,d}\, D_d(E)\, [f_d(E) - f_2(E)]$$
$$i_{2,u}(E) = (e/\hbar)\, \gamma_{2,u}\, D_u(E)\, [f_u(E) - f_2(E)] \quad (16)$$

For the scattering current caused by the demon we write (see Eq.(11))

$$i_{s,u}(E) = -i_{s,d}(E) \quad (17)$$
$$= (e/\hbar)\,(2\pi J^2 N_I)\, D_u(E)\, D_d(E)$$
$$[F_d\, f_u(E)(1-f_d(E)) - F_u\, f_d(E)(1-f_u(E))]$$

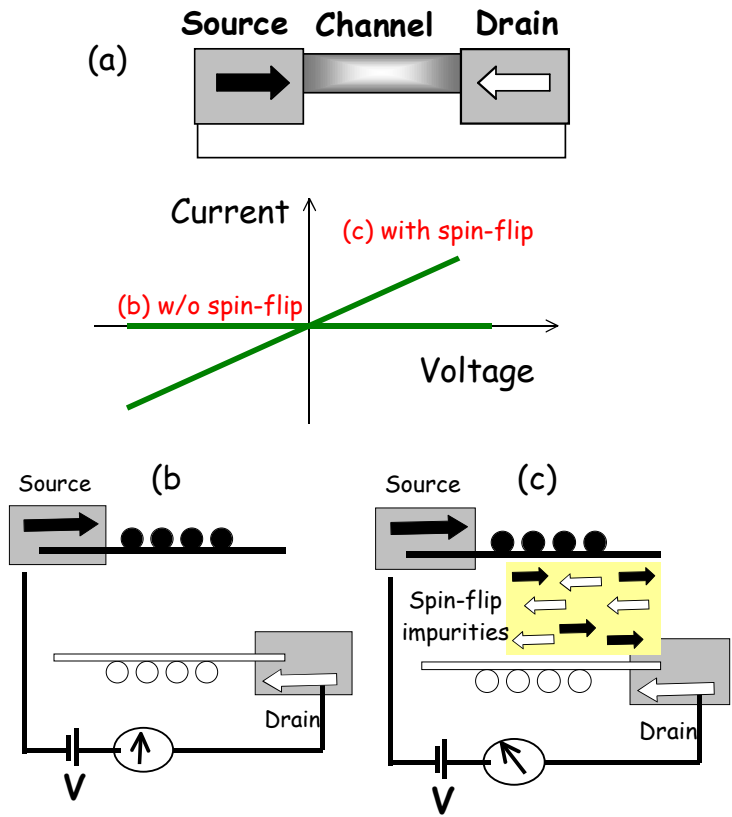

*Fig.5. Anti-parallel (AP) Spin-valve: (a) Physical structure, (b) no current flows if the contacts can discriminate between the two spins perfectly, (c) current flow is possible if impurities are present to induce spin-flip processes, Adapted from Datta, 2005b,c.*

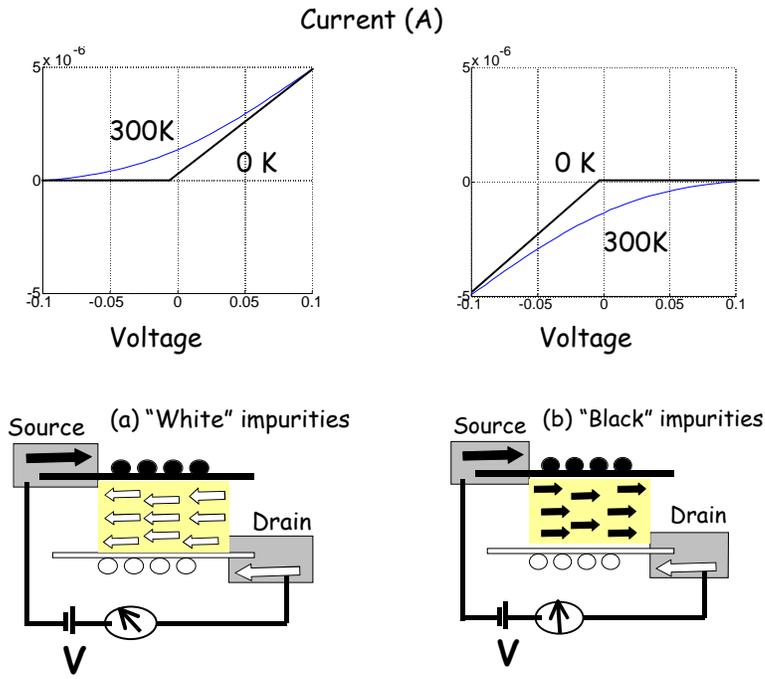

*Fig.6. Perfect AP Spin-valve with (a) white (down spin, drawn as pointing to the left) impurities and (b) black (upspin, drawn as pointing to the right) impurities. Note the non-zero current at zero voltage for non-zero temperatures. Adapted from Datta, 2005b,c.*

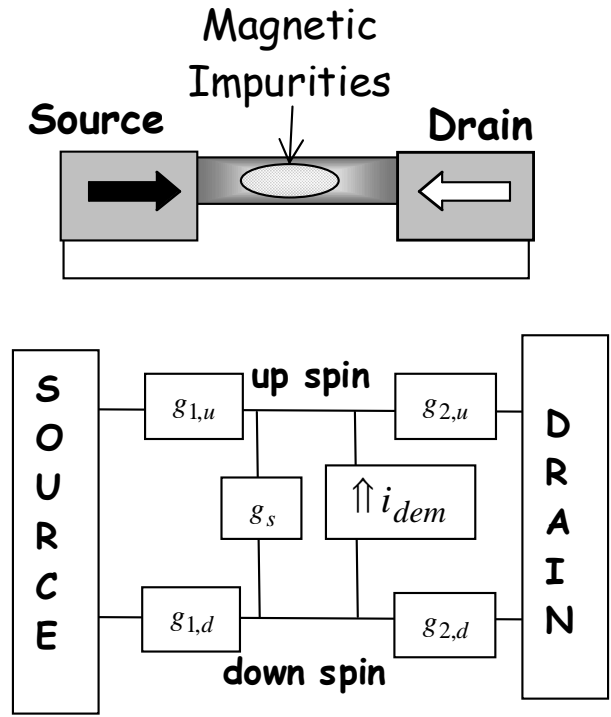

*Fig.7. An anti-parallel (AP) spin-valve with spin-flip impurities and a simple equivalent circuit to help visualize the equations we use to describe it, namely Eqs.(15) through (17), Adapted from Datta, 2005b,c.*

datta@purdue.edu

noting that we are considering an elastic demon that can neither absorb nor give up energy ($\varepsilon=0$). The first term within parenthesis in Eq.(17) represents an up electron flipping down by interacting with a down-impurity while the second term represents a down electron flipping up by interacting with an up-impurity. The strengths of the two processes are proportional to the fractions $F_d$ and $F_u$ of down and up-spin impurities. The overall strength of the interaction is governed by the number of impurities $N_I$ and the square of the matrix element J governing the electron-impurity interaction.

Eqs.(15),(16) can be visualized in terms of an equivalent circuit (Fig.7) if we think of the various f's as "voltages" since the currents are proportional to differences in 'f' just as we expect for conductances

$$g_{1,u} = (e^2/\hbar)\, \gamma_{1,u}\, D_u \qquad \text{etc.}$$

The scattering current (Eq.(17)) too could be represented with a conductance $g_s$ if we set $F_d = F_u = 0.5$

$$i_s = (e/\hbar)(\pi J^2 N_I)\, D_u(E) D_d(E)\, (f_d - f_u) \qquad (18a)$$

However, this is true only if the impurities are in equilibrium, while the interesting current-voltage characteristics shown in Fig.6 require an out-of-equilibrium demon with $F_d \neq F_u$. So we write the total scattering current from Eq.(17) as a sum of two components, one given by Eq.(18a) and another proportional to ($F_d - F_u$) which we denote with a subscript 'dem':

$$i_{dem} = (F_d - F_u)\, eN_I / \tilde{\tau}_0(E) \qquad (18b)$$

$$\frac{1}{\tilde{\tau}_0(E)} = (1/\hbar)(\pi J^2)\, D_u(E) D_d(E)$$
$$[f_d(1-f_u) + f_u(1-f_d)]$$

Eqs.(16) and (17) can be solved to obtain the distribution functions $f_u(E)$ and $f_d(E)$ by imposing the requirement of overall current conservation (cf. Eq.(11)):

$$i_{1,u}(E) = i_{2,u}(E) + i_{s,u}(E) \qquad (19)$$
$$i_{1,d}(E) = i_{2,d}(E) + i_{s,d}(E)$$

The currents are then calculated and integrated over energy to obtain the terminal currents shown in Fig.6. We can also find the energy currents using equations like Eqs.(14), (15) and the results are shown in Fig.8 for $F_d - F_u = -1$ and for $F_d - F_u = 0$ each with a voltage difference of $2\, k_B T = 50$ mV between the two terminals. With $F_d - F_u = 0$ the direction of current is in keeping with an external battery driving the device. But with $F_d - F_u = -1$, the external current flows against the terminal voltage indicating that the device is acting as a source of energy driving a load as shown in the inset. This is also borne out by the energy current flow which shows a step up at each interface indicating that *energy is being absorbed from the contacts* (~ 10 nW from each) and delivered to the external load.

But isn't this exactly what the second law forbids? After all if we could just absorb energy from our surroundings (the contacts) and do useful things, there would be no energy problem. The reason this device is able to perform this impossible feat is that the impurities are assumed to be held in a non-equilibrium state with very low entropy. A collection of $N_I$ impurities can be unpolarized in $2^{N_I}$ different ways having an entropy of $S = N_I k_B \ln 2$. But it can be completely polarized ($F_d - F_u = \pm 1$) in only one way with an entropy of $S = 0$. What this device does is to exchange entropy for energy. Many believe that the universe as a whole is evolving the same way, with constant total energy, from a particularly low entropy state continually towards a higher entropy one. But that is a different matter.

To have our device continue delivering energy indefinitely we will need an external source to maintain the impurities in their low entropy state which will cost energy. The details will depend on the actual mechanism used for the purpose but we will not go into this. Note that if we do not have such a mechanism, the current will die out as the spins get unpolarized. This depolarization process can be described by an equation of the form:

$$\frac{dP_I}{dt} + \left(\frac{1}{\tau_I} + \frac{1}{\tau_0}\right) P_I = \frac{\int dE\, i_s(E)}{eN_I} \qquad (20)$$

$$\frac{1}{\tau_0} = \int \frac{dE}{\tilde{\tau}_0(E)}$$

where $\tilde{\tau}_0$ and $i_s$ are related to the scattering current as defined in Eqs.(18), while the additional time constant $\tau_I$ represents processes unrelated to the channel electrons by which impurities can relax their spins.

### 4. INELASTIC DEMON

We have argued above that although one can extract energy from polarized impurities, energy is needed to keep them in that state since their natural high entropy state is the unpolarized one. It would seem that one way to keep the spins naturally in a polarized state is to use a nanomagnet, a collection of spins driven by a ferromagnetic interaction that keeps them all pointed in the same direction. Could such a magnet remain polarized naturally and enable us to extract energy from the contacts

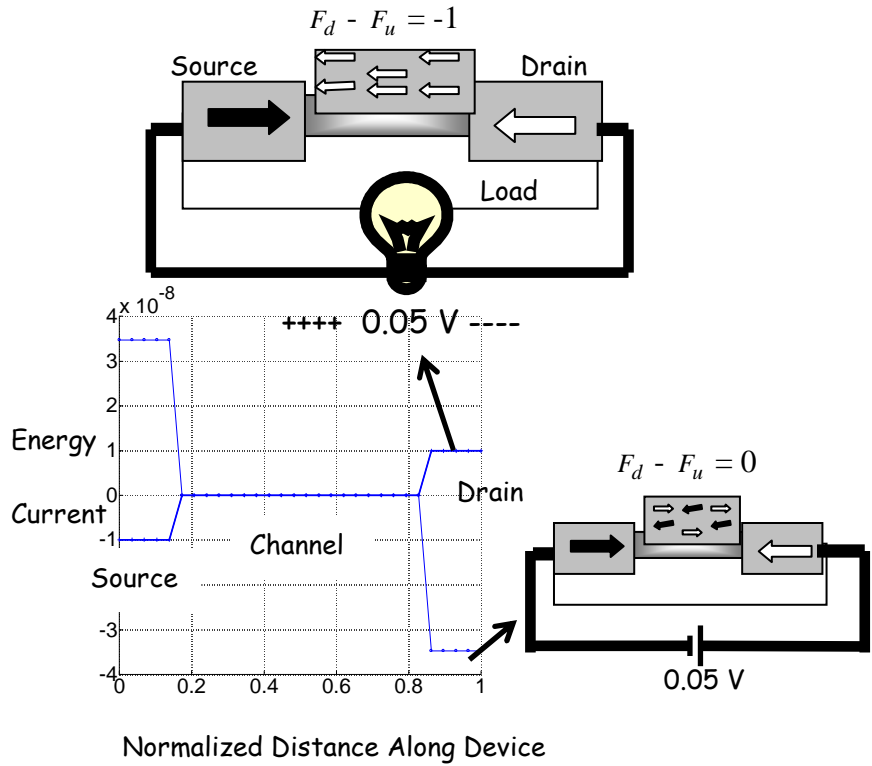

Fig.8. Energy currents with $F_d - F_u = -1$ and with $F_d - F_u = 0$ each with a voltage difference of $2 k_B T = 50$ mV between the two terminals. With $F_d - F_u = 0$ the direction of current is in keeping with an external battery driving the device. But with $F_d - F_u = -1$, the external current flows against the terminal voltage indicating that the device is acting as a source of energy driving a load as shown in the inset. This is also borne out by the energy current flow which shows a step up at each interface indicating that energy is being absorbed from the contacts (~ 10 nW from each) and delivered to the external load.

datta@purdue.edu

forever? The answer can be "yes" if the magnet is at a different temperature from the electrons. What we then have is a heat engine working between two reservoirs (the electrons and the magnet) at different temperatures and we will show that its operation is in keeping with Carnot's principle as required by the second Law.

To model the interaction of the electrons with the nanomagnet we need to modify the expression for the scattering current (Eq.(17)) for it now takes energy to flip a spin. We can write

$$i_{s,u}(E) = -i_{s,d}(E) \qquad (21)$$
$$= (e/\hbar)(2\pi J^2 N_I) \int d\varepsilon \, D_u(E) \, D_d(E+\varepsilon)$$
$$[F(\varepsilon) f_u(E)(1 - f_d(E+\varepsilon))$$
$$- F(-\varepsilon) f_d(E+\varepsilon)(1 - f_u(E))]$$

where $F(\varepsilon)$ denotes the magnon spectrum obeying the general law stated earlier (see Eq.(12)) if the magnet has a temperature $T_D$. Eq.(21) can be solved along with Eqs.(15),(16),(19) and (21) as before to obtain currents, energy currents etc. But let us first try to get some insight using simple approximations.

If we assume that the electron distribution functions $f_u(E)$ and $f_d(E)$ are described by Fermi functions with electrochemical potentials $\mu_u$ and $\mu_d$ respectively and temperature T, and make use of Eq.(12) we can rewrite the scattering current from Eq.(21) in the form

$$i_{s,u}(E) = -i_{s,d}(E)$$
$$= (e/\hbar)(2\pi J^2 N_I) \int d\varepsilon \, D_u(E) \, D_d(E+\varepsilon)$$
$$F(\varepsilon) f_u(E)(1 - f_d(E+\varepsilon))$$
$$\left(1 - \frac{F(-\varepsilon)}{F(+\varepsilon)} \frac{f_d(E+\varepsilon)}{1 - f_d(E+\varepsilon)} \frac{1 - f_u(E)}{f_u(E)}\right)$$

$$= (e/\hbar)(2\pi J^2 N_I) \qquad (22)$$
$$\int d\varepsilon \, D_u(E) \, D_d(E+\varepsilon) F(\varepsilon) f_u(E)(1 - f_d(E+\varepsilon))$$
$$\left[1 - \exp\left(\frac{\varepsilon}{k_B T_D} + \frac{\mu_d - \varepsilon - \mu_u}{k_B T}\right)\right]$$

If we further assume the exponent to be small, so that $1 - \exp(-x) \approx x$, we can write

$$i_{s,u} = -i_{s,d} \qquad (23)$$
$$\sim \left[\frac{\mu_u - \mu_d}{k_B T}\right] + \left[\frac{\varepsilon}{k_B T} - \frac{\varepsilon}{k_B T_D}\right]$$

The first term represents a dissipative current proportional to the potential difference and can be represented by a conductance like the $g_s$ in Fig.7, while the second term is the demon induced source term that can be harnessed to do external work. It vanishes when the demon temperature $T_D$ equals the electron temperature T.

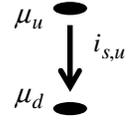

To be specific, let us assume that the demon is cooler than the rest of the device ($T_D < T$) so that

$$\frac{\varepsilon}{k_B T} - \frac{\varepsilon}{k_B T_D} < 0$$

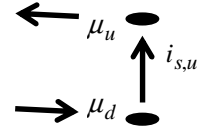

giving rise to a flow of electrons out of the upspin node and back into the downspin node. If we use this to drive an external load then $\mu_u - \mu_d > 0$, which will tend to reduce the net current given by Eq.(23) and the maximum output voltage one can get corresponds to "open circuit" conditions with zero current:

$$\frac{\mu_u - \mu_d}{k_B T} \leq \frac{\varepsilon}{k_B T_D} - \frac{\varepsilon}{k_B T} \qquad (24)$$

This expression has a simple interpretation in terms of the Carnot principle. Every time an electron flows around the circuit giving up energy $\varepsilon$ to the demon, it delivers energy $\mu_u - \mu_d$ to the load. All this energy $\mu_u - \mu_d + \varepsilon$ is absorbed from the contacts and the Carnot principle requires that

Energy from contacts / T $\leq$ Energy to demon / $T_D$

that is, $\quad \dfrac{\mu_u - \mu_d + \varepsilon}{T} \leq \dfrac{\varepsilon}{T_D}$

which is just a restatement of Eq.(24). Note that usual derivations of the Carnot principle do not put the system simultaneously in contact with two reservoirs at different temperatures as we have done. Our treatment is closer in spirit to the classic discussion of the ratchet and pawl in the Feynman Lectures [Feynman et.al. 1963].

Fig.9 shows numerical results obtained by solving Eqs.(21),(19),(15) and (16). The extracted power is a maximum (Fig.9d) when the output voltage is somewhere halfway between zero and the maximum output voltage of 80 mV (a few $k_B T$).Fig.9b shows the energy current profile at an output voltage of 50 mV: energy is absorbed from the source and drain contacts and given up partly to

*Supriyo Datta*

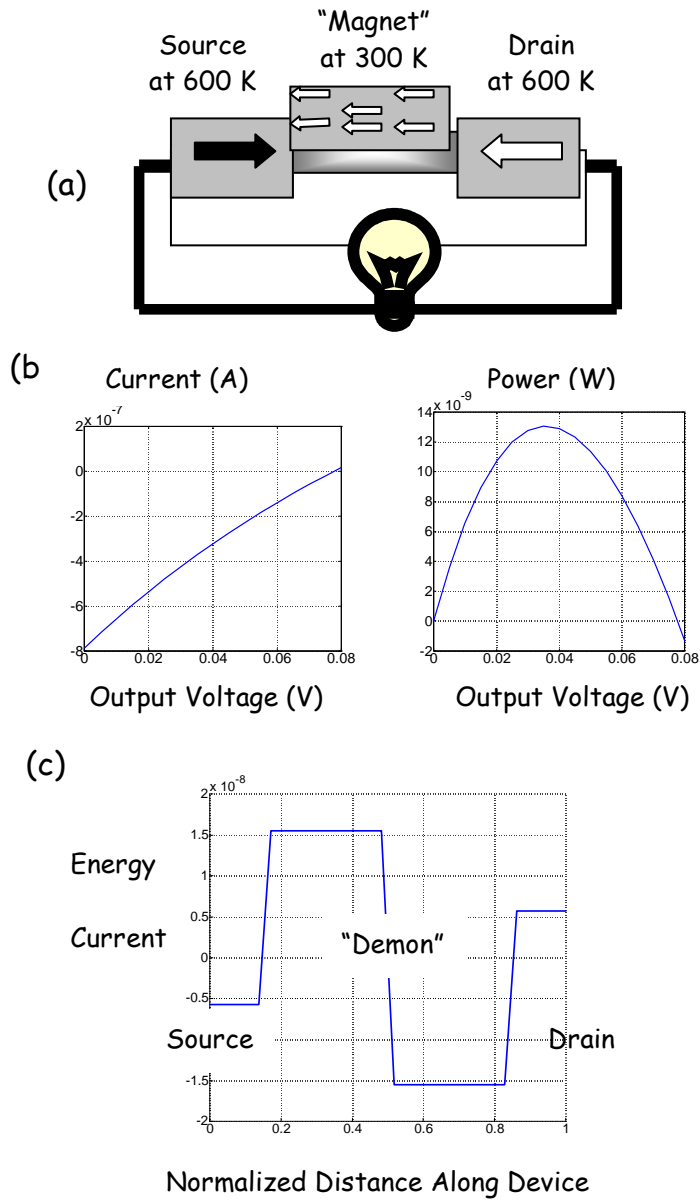

*Fig.9. Heat engine: (a) AP Spin-valve with a cooled magnet as Maxwell's demon controlling the flow of electrons. (b) Output current and output power versus output voltage as the load is varied from short circuit (V=0) to open circuit conditions (I=0). The spectrum of the magnet is assumed to consist of a single energy $\varepsilon = 2k_BT$ = 50 meV. (c) Energy current profile assuming a load such that the output voltage is 50 mV.*

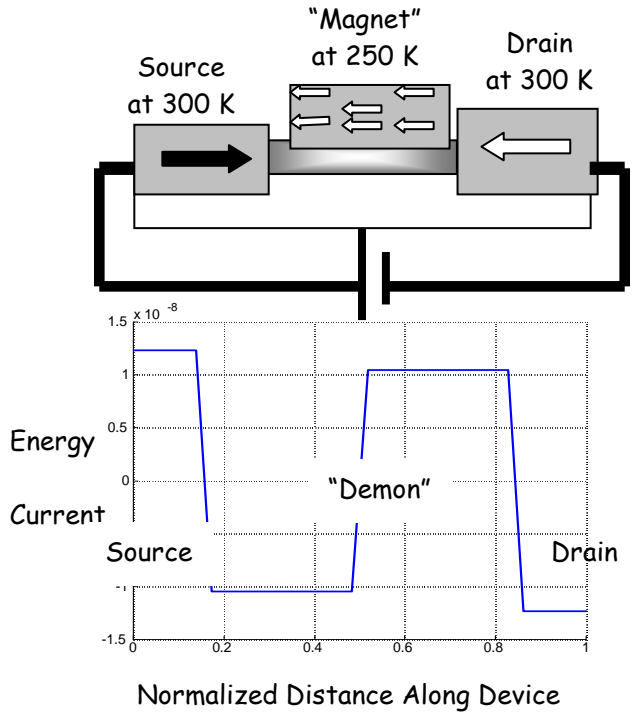

*Fig.10. (a) Heat engine from Fig.9 operated as a refrigerator by applying an external battery to inject downspin (white) electrons from the drain that flip down the thermally created upspins in the magnet, thus cooling it. (b) Energy current profile showing that energy is absorbed from the external battery and the demon and dissipated in the source and drain contacts.*

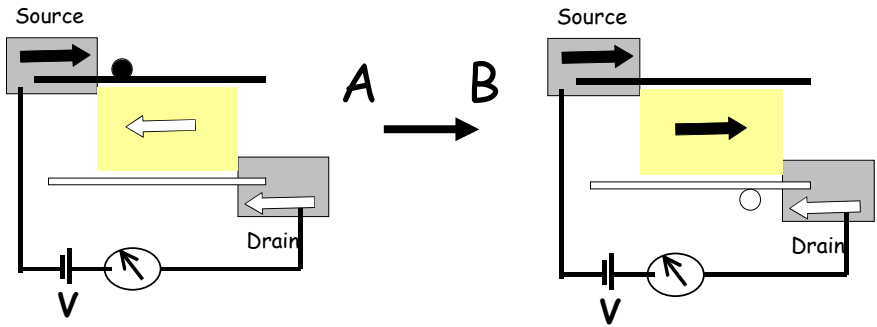

*Fig.11. We have described spin-flip processes in terms of a direct conversion from state A to state B. Quantum mechanics, however, requires an intermediate state consisting of a superposition of A and B before wavefunction collapse reduces it to a B. This state may have a significant role in devices with weak contacts and strong interactions in the channel, requiring a model that can deal with entangled states.*

*Supriyo Datta*

the demon and the rest is delivered to the external load. The efficiency (energy given to load / energy absorbed from contact) is a maximum close to open circuit conditions, but the energy delivered is very small at that point as evident from Fig.9d.

As one might expect, one can also operate the same device as a refrigerator by using an external source that seeks to inject downspins (white) from the drain contact that flip back thermally created upspins in the magnet thus cooling it. It is evident from the energy current profile shown in Fig.10 that in this case, energy is absorbed from the demon and from the battery and given up to the source and drain contacts.

## 5. ENTANGLED DEMON

In this talk I have tried to introduce a simple transparent model showing how out-of-equibrium demons suitably incorporated into nanodevices can achieve energy conversion. At the same time this model illustrates the fundamental role played by "contacts" and "demons" in these processes. I would like to end by pointing out another aspect of contacts that I believe is important in taking us to our next level of understanding. The basic point can be appreciated by considering a simple version of the spin capacitor we started with (see Fig.5) but having just one impurity (Fig.11). No current can flow in this structure without spin-flip processes since the source injects black (up) electrons while the drain only collects white (down) electrons. But if the black electron interacts with the white impurity(A) to produce a black impurity then the white electron can be collected resulting in a flow of current.

The process of conversion
    From    A: Black electron ⊗ White impurity
    To      B: White electron ⊗ Black impurity
is incorporated into our model through the scattering current (see Eq.(17)). A more complete quantum transport model involving matrices (Fig.3a) rather than numbers (Fig.3b) could be used to describe this effect, but the essential underlying assumption in either case is that the state of the electron-impurity system changes from A to B. Quantum mechanics, however, paints a different picture of the process involving an intermediate *entangled state*. It says that the system goes from A into a state consisting of a superposition of A and B and it is only when the electron is collected by the drain that the wavefunction collapses to a B. If the collection rates $\gamma_{1,2}/\hbar$ are much larger than the interaction rate per impurity $\gamma_s/\hbar N_I$, we expect the entangled state to play a minor role. But this may not be true of devices with weak contacts and strong interactions in the channel, requiring a model that can deal with entangled states.

Entangled states are difficult to describe within the conceptual framework we have been using where both the electrons and the impurity are assumed to exist in independent states. It is hard to describe a "conditional state" where the electron is black if the impurity is white or vice versa, let alone a superposition of the two. To account for this entangled state we need to treat the electron and impurity as one big system and write rate equations for it, in the spirit of the many-electron rate equations widely used to treat Coulomb blockade but the standard approach [Beenakker 1991, Likharev 1999] needs to be extended to include coherences and broadening. This is an area of active research [see for example Braun et.al. 2004, Braig and Brouwer 2005] where an adequate general approach does not yet exist.

Actually correlated states (classical version of entanglement) were an issue even before the advent of quantum mechanics. Boltzmann ignored them through his assumption of "molecular chaos" or "Stohsslansatz",and it is believed that it is precisely this assumption that leads to irreversibility [see for example, McQuarrie 1976]. An intervening entangled or correlated state is characteristic of all "channel"-"contact" interactions, classical or quantum, and the increase in entropy characteristic of irreversible processes can be associated with the destruction or neglect of the correlation and/or entanglement generated by the interaction (see for example, Zhang 2007, Datta 2007). This aspect is largely ignored in today's transport theory, just as even the presence of a contact was barely acknowledged before the advent of mesoscopic physics. But new experiments showing the effect of entanglements on current flow are on the horizon and will hopefully lead us to the next level of understanding.

This work was supported by the Office of Naval Research under Grant No. N00014-06-1-0025 and the Network for Computational Nanotechnology.

datta@purdue.edu

*Supriyo Datta*